\documentclass[12pt]{article}
\usepackage{a41}
\usepackage{epsfig}
\usepackage{axodraw}
\usepackage{amsmath}
\usepackage{amssymb}
\newcommand\e{{\rm e}}
\newcommand\PP{{\mathbb{P}}}
\newcommand\XINT{\int_{-\infty}^{+\infty}}
\newcommand\YINT{\int_{-1}^{+1}}
\newcommand{\DIRJ}{{\cal Q}^J_{\alpha}(p_2,p_1)}
\newcommand{\DIRJF}{{\cal Q}^{J,5}_{\alpha}(p_2,p_1)}
\newcommand{\DIRA}{{\cal Q}_{D \alpha}(p_2,p_1)}
\newcommand{\PAULA}{{\cal Q}_{P \alpha}(p_2,p_1)}
\newcommand{\bea}{\begin{eqnarray}}
\newcommand{\bq}{\begin{equation}}
\newcommand{\eea}{\end{eqnarray}}
\newcommand{\eq}{\end{equation}}
\newcommand{\gap}{\stackrel{>}{\sim}}
\newcommand{\lap}{\stackrel{<}{\sim}}
\newcommand{\fem}{$f_2^{em}$}
\newcommand{\lam}{$\Lambda$}
\newcommand{\qsq}{$Q^2$}
\newcommand{\alsq}{$\alpha_s(Q^2)$}
\newcommand{\dals}{$\delta \alpha_s$}
\newcommand{\pbinv}{pb^{-1}}
\newcommand{\lsim}{\raisebox{-0.07cm   }
{$\, \stackrel{<}{{\scriptstyle\sim}}\, $}}
\newcommand{\gsim}{\raisebox{-0.07cm   }
{$\, \stackrel{>}{{\scriptstyle\sim}}\, $}}
\newcommand{\dfeq}{\stackrel{\scriptsize =}{\scriptsize df}}
\newcommand{\als}{\alpha_s}
\newcommand{\aqu}{\langle Q^2 \rangle}
\newcommand\pB{[\mbox{pb}]}
\newcommand\MSbar{$\overline{\mbox{MS}}$}
\newcommand\order{{\cal O}}
\newcommand\mat{{\cal M}}
\newcommand\ds{\displaystyle}
\newcommand\MeV{\,\mbox{MeV}}
\newcommand\GeV{\,\mbox{GeV}}
\newcommand\TeV{\,\mbox{TeV}}
\newcommand\secu{\,\mbox{sec}}
\newcommand\kvec{\mbox{\boldmath $k$}}
\newcommand\pvec{\mbox{\boldmath $p$}}
\newcommand\lvec{\mbox{\boldmath $l$}}
\newcommand\MT{Morfin and Tung}
\newcommand\CALL{{\cal L}}
\newcommand\KG{\kappa_G}
\newcommand\KA{\kappa_A}
\newcommand\LG{\lambda_G}
\newcommand\LA{\lambda_A}
\newcommand\PD{\not p}
\newcommand\MP{M^2_{\Phi}}
\newcommand\BE{\beta}
\newcommand\XL{\log \left| \frac{1 + \B}{1 -\B}\right |}
\newcommand\SM{\frac{\hat{s}}{M_{\Phi}^2}}
\newcommand\SMM{\frac{\hat{s}^2}{M_{\Phi}^4}}
\newcommand\SMMM{\frac{\hat{s}^3}{M_{\Phi}^6}}
\newcommand\SMMMM{\frac{\hat{s}^4}{M_{\Phi}^8}}
\newcommand\BC{\beta^2 \cos^2 \theta}
\newcommand\CB{\beta^2 \cos^2 \theta}
\newcommand\CBB{\beta^4 \cos^4 \theta}
\newcommand\CBBB{\beta^6 \cos^6 \theta}
\newcommand\CBBBB{\beta^8 \cos^8 \theta}
\newcommand\SH{\hat{s}}
\newcommand\uh{\hat{u}}
\newcommand\thh{\hat{t}}
\newcommand\sh{\hat{s}}
\newcommand\epsi{\varepsilon}
\newcommand\ea{\varepsilon_1}
\newcommand\eb{\varepsilon_2}
\newcommand\Ea{\epsilon_1}
\newcommand\Eb{\epsilon_2}
\newcommand\ch{{\rm{ch}}}
\newcommand\shh{{\rm{sh}}}
\newcommand\thhh{{\rm{th}}}
\newcommand\st{\sin^2 \theta}
\newcommand\BR{\left(1 - \CB \right)}
\newcommand\ka{\kappa_1}
\newcommand\kb{\kappa_2}
\newcommand\kap{\kappa_1'}
\newcommand\kbp{\kappa_2'}
\newcommand\xx{\tilde{x}}
\newcommand\GG{\tilde{g}}
\newcommand\hh{\tilde{h}}
\newcommand\AAA{\alpha_1}
\newcommand\AB{\alpha_2}
\newcommand\Kvec{\mbox{\boldmath $K$}}
\newcommand\Pvec{\mbox{\boldmath $P$}}
\newcommand\Uvec{\mbox{\boldmath $U$}}
\newcommand\Vvec{\mbox{\boldmath $V$}}
\newcommand\bx{\overline{x}}
\newcommand\by{\overline{y}}
\newcommand\FFA{\mbox{$\widetilde{F}^a$}}
\newcommand\ub{\mbox{$\overline{u}(p_2,S_2)$}}
\newcommand\uu{\mbox{$u(p_1,S_1)$}}
\newcommand\g{\mbox{$\gamma$}}
\newcommand\T{{\sf T}}
\newcommand\fP{{I\!P}}
\begin{document}
\noindent
\sloppy
\thispagestyle{empty}
\begin{flushleft}
DESY 02--011 \hfill
{\tt hep-ph/0202077}\\
January  2002
\end{flushleft}
%
\vspace*{\fill}
\begin{center}
{\LARGE\bf Polarized Deep Inelastic Diffractive }

\vspace{2mm}
{\LARGE\bf $ep$ Scattering: Operator Approach}

\vspace{2cm}
\large
Johannes Bl\"umlein$^a$  and
Dieter Robaschik$^{a,b}$
\\
\vspace{2em}
\normalsize
{\it $^a$~Deutsches Elektronen--Synchrotron, DESY,\\
Platanenallee 6, D--15738 Zeuthen, Germany}
\\
\vspace{2em}
{\it $^b$~Brandenburgische Technische Universit\"at Cottbus, 
Fakult\"at 1,}\\
{\it  PF 101344, D--03013  Cottbus, Germany} \\
\end{center}
\vspace*{\fill}
%
\begin{abstract}
\noindent
Polarized inclusive deep--inelastic diffractive scattering is dealt with
in a quantum field theoretic approach. The process can be described in 
the general framework of non--forward scattering processes using the 
light--cone expansion in the generalized Bjorken region applying the
generalized optical theorem. The diffractive structure functions 
$g_1^{D(3)}$ and $g_2^{D(3)}$ are calculated in the twist--2 approximation 
and are expressed by diffractive parton distributions, which are derived 
from pseudo-scalar two--variable operator expectation values. In this
approximation the structure functions $g_2^{D(3)}$ is obtained from 
$g_1^{D(3)}$ by a   Wandzura--Wilczek relation similar as for deep
inelastic scattering. The evolution equations are given. We also comment 
on the higher twist contributions in the light--cone expansion.
\end{abstract}
\vspace*{\fill}
\newpage
\section{Introduction}
\label{sec-1}

\vspace{1mm}
\noindent
Unpolarized deep inelastic diffractive lepton--nucleon scattering was 
observed at the electron--proton collider HERA some years ago~\cite{EXP1}.
In the region of hard diffractive scattering this process is described 
by structure functions which are represented by diffractive parton 
distributions. They depend on two scaling variables
$x$ and $x_{\PP}$ and are different from the parton densities of deep 
inelastic scattering. New diffractive parton densities are expected
to occur in polarized deep inelastic diffractive lepton nucleon 
scattering. They can be measured at potential future polarized $ep$
facilities capable to probe the kinematic range of small $x$~, 
c.f.~\cite{SXPOL}. Dedicated future experimental studies of this process
can reveal the helicity structure of the non--perturbative color--neutral 
exchange of diffractive scattering with respect to the quark and
gluon structure and how the nucleon spin is viewed under a diffractive
exchange.
At short distances the problem can be clearly 
separated into
a part, which can be described within perturbative QCD, and another
part which is thoroughly non--perturbative. In this paper we use the
light--cone expansion to describe the process of polarized diffractive
deep--inelastic scattering similar to a recent study for the unpolarized
case~\cite{BRDIFF}. While the scaling violations of the process can be
calculated within perturbative QCD, the polarized diffractive 
two--variable parton densities are non--perturbative and can be related 
to expectation values of (non--)local operators. Their Mellin-moments 
with respect to the variable $\beta = x/x_{\PP}$ may, in principle, be 
calculated on the lattice and one may try to understand the ratios of 
these moments and those for the related deep--inelastic process w.r.t. 
their scaling violations as being measurable in future experiments.

In this paper we describe the process of polarized deep--inelastic
diffractive scattering, which is a non--forward process in its hadronic
variables, at large space--like momentum transfer $q^2$. In this approach
there is no need to refer to any specific mechanism of color--singlet
exchange. It is completely sufficient to select the process by a 
rapidity gap between the final state proton and the other diffractively  
produced hadrons, which is sufficiently large. 
The operator formulation allows straightforwardly the
description of also higher twist operators in the light cone expansion,
which is potentially more involved in other scenarios~\cite{FACT},
to which we agree on the level of twist--2.

We firstly derive the  Lorentz--structure of the process for the
general kinematics, before we specify to the case of 
$-t = -(p_2 - p_1)^2, M^2 << -q^2$ which is
often met in experiment. The 
diffractive parton densities are derived on the level of the twist--2 
operators. In this approximation the scattering cross sections are 
described by two structure functions $g_1^{D(3)}(x,Q^2,x_\PP)$ and
$g_2^{D(3)}(x,Q^2,x_\PP)$ for pure electromagnetic scattering\footnote{
The exchange of electro--weak gauge bosons requires at least five 
structure functions \cite{BK}. QED radiative corrections to the process 
were given in \cite{PRAD}.}. Also in the present case it turns out that 
the structure functions are related by the Wandzura--Wilczek 
relation~\cite{WW}. Analogously to the unpolarized case, 
Ref.~\cite{BRDIFF}, the anomalous dimensions ruling the evolution of
the polarized diffractive parton densities turn out to be those for
deep--inelastic forward scattering.

\section{Lorentz Structure}
\vspace*{-5mm}\noindent
\label{sec-2}

\vspace{1mm}
\noindent
The process of deep--inelastic diffractive scattering is described by the
diagram Figure~1.
The differential scattering cross section for single--photon 
exchange is given by
\begin{equation}
\label{eqD1}
d^5 \sigma_{\rm diffr}
= \frac{1}{2(s-M^2)} \frac{1}{4} dPS^{(3)} \sum_{\rm spins}
\frac{e^4}{Q^2} L_{\mu\nu} W^{\mu\nu}~.
\end{equation}
Here $s=(p_1+l)^2$ is the cms energy of the process squared and $M$ 
denotes the nucleon mass.

\vspace*{0.8cm}
\begin{center}
\begin{picture}(-50,100)(0,0)
\setlength{\unitlength}{0.2mm}
\SetWidth{1.5}
\ArrowLine(-150,100)(-100,100)
\ArrowLine(-100,100)(-50,120)
\Photon(-100,100)(-70,70){5}{5}%
\ArrowLine(-30,30)(0,0)
\ArrowLine(-100,0)(-70,30)
\SetWidth{5}
\ArrowLine(-30,70)(0,100)
\SetWidth{1.5}
\CCirc(-50,50){26}{Black}{Yellow}
\setlength{\unitlength}{1pt}
\Text(-160,110)[]{$l$}
\Text(-40,130)[]{$l'$}
\Text(-90,75)[]{$q$}
\Text(-110,-10)[]{$p_1$}
\Text(10,-10)[]{$p_2$}
\Text(10,110)[]{$M_X$}
\end{picture}
\end{center}

\vspace*{1mm}
\noindent
\begin{center}
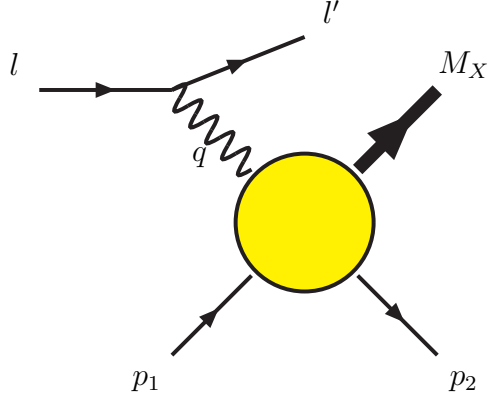
{\sf Figure~1:~The virtual photon-hadron amplitude for 
diffractive $ep$ scattering} 
\end{center}
The phase space $dPS^{(3)}$ depends on five variables
since one 
final state mass varies. They can be chosen as 
Bjorken~$x= Q^2/(W^2+Q^2-M^2)$, the photon virtuality $Q^2=-q^2$, 
$t = (p_1 - p_2)^2$, a variable describing the non--forwardness w.r.t.
the incoming proton direction,
\begin{equation}
\label{eqV1}
x_{\PP} = - \frac{2\eta}{1-\eta} = 
\frac{Q^2 + M_X^2 - t}{Q^2 + W^2 -M^2} \geq  x~,
\end{equation}
demanding $M_X^2 > t$ and where
\begin{equation}
\label{eqV2}
\eta = \frac{q.(p_2 - p_1)}{q.(p_2+p_1)}~\epsilon~\left[-1,\frac{-x}{2-x}
\right]~,
\end{equation}
and $\Phi$ the angle between the lepton plane $\pvec_1 \times \lvec$ and
the hadron plane $\pvec_1 \times \pvec_2$,
\begin{equation}
\label{eqV3}
\cos \Phi = \frac{(\pvec_1 \times \lvec).(\pvec_1 \times  \pvec_2)}
                 {|\pvec_1 \times \lvec ||\pvec_1 \times  \pvec_2|}~.
\end{equation}
$W^2 = (p_1+q)^2$ and  $M_X^2 = (p_1+q-p_2)^2$ denote the hadronic
mass squared and the square of the diffractive mass, respectively.
The process of hard diffractive scattering is characterized by a large
rapidity--gap of the order $\Delta y \sim \ln(1/x_{\PP})$ \cite{COL}.
As we will show below it is {\it this property}, which is sufficient for 
our treatment below and no reference to a special kind of a 
non--perturbative color--neutral exchange is not needed.\footnote{
Indeed, the literature offers a large host of different pomeron 
models, c.f.~\cite{POM},
to describe these processes. The fact that many of the descriptions yield
similar results at equally large rapidity gaps and the same kinematic 
variables supports our observation.}

Unpolarized deep inelastic diffractive scattering was considered in a
previous paper~\cite{BRDIFF} in detail. Here we focus on the polarized 
part only, which can be measured in terms of a polarization asymmetry
\begin{equation}
\label{eqP1}
A(x,Q^2,x_{\PP},S_\mu) = \frac{d^5 \sigma(S_{\mu}) - d^5 \sigma(-S_\mu)}
                              {d^5 \sigma(S_{\mu}) + d^5 \sigma(-S_\mu)}~.
\end{equation}
$S_\mu$ is the spin vector of the incoming proton with $S = S_1$ and
$S.p_1 =0$.
Since the cross sections are linear functions in the initial--state
state proton spin--vector, the denominator projects on the even and the 
numerator on the odd part in $S_\mu$. 

We consider the case of single photon exchange, which is projected by the
polarized contribution
\begin{equation}
\label{eqPLE}
L_{\mu\nu}^{\rm pol} = 2i \varepsilon_{\mu\nu\rho\sigma} l^\rho q^\sigma
\end{equation}
to the leptonic tensor. Since the electromagnetic current is conserved,
the strong interactions conserve parity and are even under 
time--reversal,\footnote{Here we disregard potential contributions due to
strong CP--violation~\cite{THOOFT}, because of the smallness of the 
$\theta$--parameter, $|\theta| < 3 \cdot 10^{-9}$~\cite{THETA}.} 
and the hadronic tensor has to be hermitic due to Eq.~(\ref{eqPLE}), the
following relations hold~\cite{TM}~:
\begin{alignat}{3}
\label{eqP2A}
{\sf Current~conservation:}&~~q^{\mu}~W_{\mu\nu}(q,p_1,S_1,p_2,S_2)
  &=&
W_{\mu\nu}(q,p_1,S_1,p_2,S_2)~q^{\nu} = 0~,\\
{\sf P~~invariance:}&~~W_{\mu\nu}(\overline{q},%
\overline{p}_1,-\overline{S}_1,\overline{p}_2,%
-\overline{S}_2)  &=&
W^{\mu\nu}(q,p_1,S_1,p_2,S_2)~, \\
\label{eqP2C}
{\sf T~~invariance:}&~~W_{\mu\nu}(\overline{q},%
\overline{p}_1,\overline{S}_1,\overline{p}_2,\overline{S}_2)   &=&
\left[W^{\mu\nu}(q,p_1,S_1,p_2,S_2)\right]^*, \\
\label{eqP2B}
{\sf Hermiticity:}&~~W_{\mu\nu}(q,p_1,S_1,p_2,S_2)
 &=&
\left[W_{\nu\mu}(q,p_1,S_1,p_2,S_2)\right]^*,
\end{alignat}
with $\overline{a}_\mu = a^{\mu}$. Constructing the hadronic tensor we 
seek a structure which is linear in the initial proton spin. 
Upon noting that
\begin{eqnarray}
\label{eqP3}
\varepsilon^{\mu\nu\alpha\beta} = - \varepsilon_{\mu\nu\alpha\beta}
\end{eqnarray}
the spin  pseudovector  $S_{1\mu}$ has to occur together
with the Levi--Civita pseudo--tensor.
The most general asymmetric  hadronic tensor, which
obeys Eqs.~(\ref{eqP2A}--\ref{eqP2B}),
is\footnote{A sub-set of this structure based
on $p,q$ and $S$ was considered in Ref.~\cite{DJT}.}
\begin{alignat}{10}
\label{eqH2}
W_{\mu\nu} &=&~~
i \left[ \hat{p}_{1\mu} \hat{p}_{2\nu}-
\hat{p}_{1\nu} \hat{p}_{2\mu} \right] \varepsilon_{p_1,p_2,q,S} 
&\frac{W_1}{M^6}&~~
                &+&~~
i \left[ \hat{p}_{1\mu} \varepsilon_{\nu S p_1 q}
     - \hat{p}_{1\nu} \varepsilon_{\mu S p_1 q} \right] 
&\frac{W_2}{M^4}&
\nonumber \\ &+&~~
i \left[ \hat{p}_{2\mu} \varepsilon_{\nu S p_1 q}
     - \hat{p}_{2\nu} \varepsilon_{\mu S p_1 q} \right]
&\frac{W_3}{M^4}&~~
                &+&~~
i \left[ \hat{p}_{1\mu} \varepsilon_{\nu S p_2 q}
     - \hat{p}_{1\nu} \varepsilon_{\mu S p_2 q} \right]
&\frac{W_4}{M^4}&
\nonumber \\ &+&~~
i \left[ \hat{p}_{2\mu} \varepsilon_{\nu S p_2 q}
     - \hat{p}_{2\nu} \varepsilon_{\mu S p_2 q} \right]
&\frac{W_5}{M^4}&~~
                &+&~~
i \left[ \hat{p}_{1\mu} \hat{\varepsilon}_{\nu p_1 p_2 S}
     - \hat{p}_{1\nu} \hat{\varepsilon}_{\mu p_1 p_2 S} \right]
&\frac{W_6}{M^4}&
\nonumber\\ &+&~~ 
i \left[ \hat{p}_{2\mu} \hat{\varepsilon}_{\nu p_1 p_2 S}
     - \hat{p}_{2\nu} \hat{\varepsilon}_{\mu p_1 p_2 S} \right]
&\frac{W_7}{M^4}&~~
                &+&~~  i \varepsilon_{\mu \nu q S}
&\frac{W_8}{M^2}&~.
\end{alignat}
It is constructed out of the four--vectors $q,p_1,p_2$ and $S =S_1$.
Terms with a genuine structure $\propto M^2/q^2$ are not considered.
Here we use the abbreviations
\begin{eqnarray}
\hat{V}_\mu &=& V_\mu - q_\mu \frac{q.V}{q^2}~, \\
\hat{\varepsilon}_{\mu v_1 v_2 v_3}            &=&
    {\varepsilon}_{\mu v_1 v_2 v_3}            -
    {\varepsilon}_{q v_1 v_2 v_3} \frac{q_\mu}{q^2}~,  \\
\tilde{\varepsilon}_{\mu \nu v_1 v_2}            &=&
    {\varepsilon}_{\mu \nu v_1 v_2}            -
    {\varepsilon}_{q \nu v_1 v_2} \frac{q_\mu}{q^2}
  - {\varepsilon}_{\mu q v_1 v_2} \frac{q_\nu}{q^2}~.
\end{eqnarray}
The Schouten--relation~\cite{SCHOUT} in either of the forms
\begin{eqnarray}
\label{eqP6}
X_\mu \varepsilon_{\nu\rho\sigma\tau} &=&
X_\nu \varepsilon_{\mu\rho\sigma\tau} +
X_\rho \varepsilon_{\nu\mu\sigma\tau} +
X_\sigma \varepsilon_{\nu\rho\mu\tau} +  
X_\tau \varepsilon_{\nu\rho\sigma\mu}  \\
g_{\lambda\mu} \varepsilon_{\nu\rho\sigma\tau} &=&
g_{\lambda\nu} \varepsilon_{\mu\rho\sigma\tau} +
g_{\lambda\rho} \varepsilon_{\nu\mu\sigma\tau} +
g_{\lambda\sigma} \varepsilon_{\nu\rho\mu\tau} +  
g_{\lambda\tau} \varepsilon_{\nu\rho\sigma\mu}
\end{eqnarray}
is  used to eliminate other possible structures. Particularly, the
spin vector $S_\mu$ may always be contracted with the Levi--Civita
symbol, along with it it has to occur due to parity conservation.
Because $S.p_1 = 0$ two other structures are eliminated using
\begin{eqnarray}
q.p_1 \tilde{\varepsilon}_{\mu \nu S p_1}
&=& p_1.p_1 \varepsilon_{\nu \mu q S}
- \left[\hat{p}_{1\mu}\varepsilon_{\nu p_1 q S} - \hat{p}_{1\nu}
\varepsilon_{\mu p_1 q S}\right]\\
q.p_1 \tilde{\varepsilon}_{\mu \nu S p_2}
&=& p_1.p_2 \varepsilon_{\nu \mu q S}
- \left[\hat{p}_{1\mu}\varepsilon_{\nu p_2 q S} - \hat{p}_{1\nu}
\varepsilon_{\mu p_2 q S}\right]~.
\end{eqnarray}
The structure functions $W_i$ are real functions and are given by
\begin{equation}
\label{eqD3}
W_i = W_i(x,Q^2,x_{\PP},t)~.
\end{equation}

Let us consider the limit in which target masses can be neglected
and $t$ is very small. In this case the proton momenta become
proportional: $p_2 = z p_1$ with,
\begin{equation}
\label{eqD4}
z = 1 - x_{\PP} =  \frac{1 + \eta}{1 - \eta}~.
\end{equation}
Correspondingly the hadronic tensor simplifies to
\begin{eqnarray}
\label{eqD5}
W_{\mu\nu} &=& i \varepsilon_{\mu \nu q S} \frac{W_8}{M^2}
+ i \left[\hat{p}_{1\mu} \varepsilon_{\nu S p_1 q}
-       \hat{p}_{1\nu} \varepsilon_{\mu S p_1 q}\right] \frac{W_9}{M^4}~,
\end{eqnarray}
and contains only two structure functions,
where
\begin{eqnarray}
\label{eqD6}
W_9 = W_2 + (1-x_{\PP}) \left[W_3 + W_4 \right]
+ (1-x_{\PP})^2 W_5~.
\end{eqnarray}
One may wish to re-write Eq.~(\ref{eqD5}) further into the form which is
similar to that used in polarized deep--inelastic scattering.
\begin{eqnarray}
\label{eqD5A}
W_{\mu\nu} &=& i \varepsilon_{\mu \nu \lambda \sigma} \frac{q^\lambda
S^\sigma}{p_1.q}
g_1(x,Q^2,x_{\PP})
+ i  \varepsilon_{\mu \nu \lambda \sigma}
\frac{q^{\lambda}(p_1.q S^\sigma - S.q p_1^\sigma)}{(p_1.q)^2}
g_2(x,Q^2,x_{\PP})~.
\end{eqnarray}
This again is achieved by using the Schouten relation~Eq.~(\ref{eqP6})
noting that 
\begin{eqnarray}
\label{eqSCH1}
\hat{p}_{1\mu} \varepsilon_{\nu S p_1 q} - \hat{p}_{1\nu}
\varepsilon_{\mu S p_1 q} = - \frac{(S.q)(q.p_1)}{q^2}
\varepsilon_{\mu \nu q p_1} + \frac{(q.p_1)^2}{q^2} 
\varepsilon_{\mu \nu q S}~.
\end{eqnarray}
The relation between the structure functions
$W_{8,9}$ and $g_{1,2}$ is~:
\begin{eqnarray}
\label{eqg1}
g_1 &=& \frac{q.p_1}{M^2} W_8 \\
\label{eqg2}
g_2 &=& \frac{(q.p_1)^3}{q^2 M^4} W_9
\end{eqnarray}

Due to the dependence of the structure functions on $x_{\PP}$ or 
$\eta$,~Eq.~(\ref{eqV1}), the process is {\it non--forward} w.r.t. the
protons, although the algebraic structure of the hadronic tensor
is the same as in the forward case. Finally the generalized 
Bjorken--limit is carried out, 
\begin{eqnarray}
\label{eqBL}
2 p_1.q = 2M \nu \rightarrow \infty,~~~~~p_2.q \rightarrow \infty,~~~~~
Q^2 \rightarrow \infty~~~{\rm with}~~x~~{\rm and}~~x_{\PP} =~{\rm fixed}~,
\end{eqnarray}
which leads to (\ref{eqD5A}) using (\ref{eqg1},\ref{eqg2}).
For the scattering cross sections we consider the cases of longitudinal 
and transverse target polarization for which the initial state hadron
spin vectors are given by
\begin{eqnarray}
\label{eqSPI}
S_{\parallel} &=& (0,0,0,M)\\
S_{\perp} &=& M(0,\cos\gamma,\sin\gamma,0)~,
\end{eqnarray}
and $\gamma$ the spin direction in the plane orthogonal to the 3--momentum
$\vec{p}_1$.
In the limit $p_2 = zp_1$ and $M^2,t = 0$ the $\Phi-$integral becomes 
trivial in the case of longitudinal nucleon polarization, while it is
kept as differential variable for transverse polarization.
\begin{eqnarray}
\label{eqD7}
\frac{d^3 \sigma_{\rm diffr}(\lambda,\pm S_{\parallel})}
{dx dQ^2 d x_{\PP}} &=&  \mp 4 \pi s \lambda
\frac{\alpha^2}{Q^4} \Biggl[
y \left(2 - y - \frac{2xyM^2}{s}\right) x g_1(x,Q^2,x_\PP)
\nonumber\\ & &~~~~~~~~~~~~~~~~~~~~~~~~~~~~~~~~
- 4 x y \frac{M^2}{s} g_2(x,Q^2,x_\PP) \Biggr]   \\
\frac{d^4 \sigma_{\rm diffr}(\lambda,\pm S_{\perp})}
{dx dQ^2 d \Phi d x_{\PP}} &=& \mp 4  s \lambda \sqrt{\frac{M^2}{s}}
\frac{\alpha^2}{Q^4}
\sqrt{xy\left[1-y-\frac{xyM^2}{s}\right]} \cos(\gamma - \Phi)
\nonumber\\ & &~~~~~~~~~~~~~~~~~~~~\times
\left[yx g_1(x,Q^2,x_\PP) + 2x g_2(x,Q^2,x_\PP)\right]~,
\end{eqnarray}
where $y = q.p_1/l.p_1$ and $\lambda$ denotes the degree of longitudinal
lepton polarization.\footnote{In the case of longitudinal nucleon
polarization polarized diffractive scattering was discussed neglecting
the contribution due to the structure function $g_2$ in \cite{BARY}.}
\section{The Compton Amplitude}
\label{sec-3}

\vspace{1mm}
\noindent
We first consider the operator given by the renormalized and 
time--ordered product of two electromagnetic currents
\begin{eqnarray}
\widehat{T}_{\mu\nu}(x) &=&
RT \left[J_\mu\left(\frac{x}{2}\right)J_\nu\left(-\frac{x}{2}\right) S 
\right] \nonumber\\
&=&
 -e^2 \frac{\tilde x^\lambda}{2 \pi^2 (x^2-i\epsilon)^2}
 RT
 \left[
\overline{\psi}
\left(\frac{\tilde x}{2}\right)
\gamma^\mu \gamma^\lambda \gamma^\nu \psi
\left(-\frac{\tilde x}{2}\right)
- \overline{\psi}
\left(-\frac{\tilde x}{2}\right)
\gamma^\mu \gamma^\lambda \gamma^\nu \psi
\left(\frac{\tilde x}{2}\right)
\right] S
\end{eqnarray}
Here, $\tilde x$ denotes a light--like vector corresponding to $x$,
\begin{eqnarray}
\label{xtil}
\tilde x = x + \frac{\zeta}{\zeta^2}\left[ \sqrt{x.\zeta^2 - x^2 \zeta^2}
- x.\zeta\right]~,
\end{eqnarray}
and $\zeta$ is a subsidiary vector. Following Refs.~\cite{BGR,BR} the
operator $\widehat{T}_{\mu\nu}$ can be expressed in terms of a vector
and an axial--vector operator decomposing
\begin{eqnarray}
 \gamma_\mu \gamma_\lambda \gamma_\nu = \left[g_{\mu\lambda} g_{\nu\rho}
 + g_{\nu\lambda} g_{\mu\rho} - g_{\mu\nu} g_{\lambda\rho} \right]
 \gamma^\rho - i \varepsilon_{\mu\nu\lambda\rho} \gamma^5 \gamma^\rho~.
\end{eqnarray}
    We will consider only the contribution of the latter one, since this
yields the polarized part,
\begin{eqnarray}
 \widehat{T}_{\mu\nu}^{\rm pol}(x)  =
  i e^2 \frac{\tilde x^\lambda}{2 \pi^2 (x^2-i\epsilon)^2}
  \varepsilon_{\mu\nu\lambda\sigma}
O_5^\sigma  \left(\frac{\tilde x}{2}, -\frac{\tilde x}{2}\right)~,
\end{eqnarray}
with $\varepsilon_{\mu\nu\lambda\sigma}$ the Levi--Civita symbol.
The  bilocal axial--vector light--ray operator is
\begin{eqnarray}
\label{oo5}
O^{\alpha}_5\left(\frac{\tilde x}{2},-\frac{\tilde x}{2}\right)
&=&
\frac{i}{2} RT
\left[\overline{\psi}\left(\frac{\tilde x}{2}\right)
\gamma_5\gamma^\alpha\psi\left(-\frac{\tilde x}{2}\right)
+ \overline{\psi}\left(-\frac{\tilde x}{2}\right)
\gamma_5\gamma^\alpha\psi\left(\frac{\tilde x}{2}\right)\right]S~.
\end{eqnarray}
The polarized part of the Compton operator 
$\widehat{T}^{\rm pol}_{\mu\nu}$ is related to the diffractive scattering
cross section using Mueller's generalized optical theorem~\cite{AHM} 
(Figure~2), which moves the final state proton into an initial state 
anti-proton.

\vspace*{7mm}
\begin{center}
\begin{picture}(200,100)(0,0)
\setlength{\unitlength}{0.2mm}
\SetWidth{1.5}
\Line(-107,102)(-107,-2)
\Line(7,102)(7,-2)
\Photon(-100,100)(-70,70){5}{5}
\ArrowLine(-30,30)(0,0)
\ArrowLine(-100,0)(-70,30)
\SetWidth{5}
\ArrowLine(-30,70)(0,100)
\SetWidth{1.5}
\CCirc(-50,50){26}{Black}{Yellow}
\setlength{\unitlength}{1pt}
\Text(15,107)[]{\large $2$}
\Text(50,50)[]{\large $=~~{\rm Disc}$}
\Text(-105,115)[]{$q$}
\Text(-105,-15)[]{$p_1$}
\Text(5,-15)[]{$p_2$}
\Text(75,115)[]{$q$}
\Text(75,-15)[]{$p_1$}
\Text(95,-15)[]{$p_2$}
\Text(60,35)[]{$X$}
\Text(175,20)[]{$X$}
\Text(275,115)[]{$q$}
\Text(275,-15)[]{$p_1$}
\Text(255,-15)[]{$p_2$}
\setlength{\unitlength}{0.2mm}
\SetWidth{1.5}
\Photon(80,100)(110,70){5}{5}
\ArrowLine(80,0)(110,30)
\ArrowLine(130,30)(100,0)
\Line(150,50)(200,50)
\Line(150,45)(200,45)
\Line(150,55)(200,55)
\Line(150,40)(200,40)
\Line(150,60)(200,60)
\ArrowLine(240,30)(270,0)
\ArrowLine(250,0)(220,30)
\CCirc(130,50){26}{Black}{Yellow}
\CCirc(220,50){26}{Black}{Yellow}
\Photon(240,70)(270,100){5}{5}
\end{picture}
\end{center}

\vspace*{10mm}\noindent
\begin{center}
{\sf Figure~2:~A. Mueller's optical theorem.}
\end{center}
The polarized part of the Compton amplitude is obtained as the 
expectation value
\begin{eqnarray}
T^{\rm pol}_{\mu\nu}(x) 
= \langle p_1,S_1,-p_2,S_2|\widehat{T}_{\mu\nu}|p_1,S_1,-p_2,S_2\rangle~,
\end{eqnarray}
which is   forward w.r.t. to the direction defined by the state
$\langle p_1,-p_2|$.
The twist--2 contributions to the expectation values of the operator
(\ref{oo5})  is obtained
\begin{eqnarray}
\label{eqSCA}
\lefteqn{\hspace*{-13.5cm}
\langle p_1,S_1,-p_2,S_2|O^{A,\mu}_{5}(\kappa_+ \xx,\kappa_-\xx)|
p_1,S_1,-p_2,S_2\rangle = } \nonumber\\ \left.  \hspace*{3cm}
\int_0^1 d\lambda
\partial^{\mu}_x
\langle p_1,S_1,-p_2,S_2|O_{5}^A(\lambda \kappa_+ x,\lambda \kappa_-x)
|p_1,S_1,-p_2,S_2\rangle \right|_{x=\xx}
\end{eqnarray}
as partial derivative  of the expectation values of
\begin{eqnarray}
O^A_5(\kappa_+x,\kappa_-x)
= x^\alpha O^A_{5,\alpha}(\kappa_+x,\kappa_-x)~,
\end{eqnarray}
the corresponding            pseudo-scalar operator.  The index $A = q,G$
labels the quark-- or gluon operators, cf.~\cite{BGR}. From now on we
keep only the spin vector of the initial--state proton and sum over
that of the final--state proton.

The pseudo-scalar twist--2 quark operator matrix element has the 
following representation\footnote{For parameterizations
  of similar hadronic
matrix elements see e.g.~\cite{PARA}.} due to the overall symmetry in $x$
\begin{eqnarray}
\label{eqSCAM}
\langle p_1,S_1,-p_2|O^q(\kappa_+ x,\kappa_- x)|p_1,S_1,-p_2
\rangle =
xS
\!\!\!         \int \!\!\!
Dz ~\e^{-i \kappa_- x p_z} {f}^q_5(z_+,z_-)
\end{eqnarray}
with $S \equiv S_1$, $\kappa_- =1/2$ and
where all the trace--terms were subtracted, see
\cite{BGR,TRAC}.
${f}^A_5(z_+,z_-)$ denotes                              the scalar
two--variable distribution amplitudes and the measure $Dz$ is
\begin{eqnarray}
\label{Dz}
Dz = d z_+ d z_-
\theta(1 + z_+ + z_-) \theta(1 + z_+ - z_-)
\theta(1 - z_+ + z_-) \theta(1 - z_+ - z_-)~.
\end{eqnarray}
Here, we decomposed the vector $p_z$ as
\begin{eqnarray}
p_z = p_- z_- + p_+ z_+ =  p_- \vartheta + \pi_- z_+~,
\end{eqnarray}
with $z_{1,2}$ momentum fractions along $p_{1,2}$ and $p_{\pm} = p_2 \pm
p_1$, $z_\pm = (z_2 \pm z_1)/2$ and
\begin{eqnarray}
\vartheta = z_- + \frac{1}{\eta} z_+,~~~~~~~~~~~~~\pi_-
= p_+ - \frac{1}{\eta} p_-~,
\end{eqnarray}
with       $q.\pi_- \equiv 0$. In the limit         $M^2, t \sim 0$, in
which we work from now on, the vector $\pi_-$ even vanishes.

The Fourier--transform of the Compton amplitude is given by~\cite{BR}
\begin{eqnarray}
T_{\mu\nu}^{\rm pol}(p_1,p_2,S,q)
&=&   \int d^4x \e^{iqx} T_{\mu\nu}(x) \nonumber\\
&=& 4i \varepsilon_{\mu\nu\lambda\sigma} \int Dz \left[
\frac{Q_z^{\lambda} S^{\sigma}}{Q^2_z + i\varepsilon} - \frac{1}{2}
\frac{p_z^{\sigma} S^{\lambda}}{Q^2_z + i\varepsilon}%
+ \frac{Q_z.S}{(Q^2_z+i\varepsilon)^2}
p_z^{\sigma} Q_z^{\lambda}\right] F_5(z_+,z_-),
\end{eqnarray}
with $Q_z = q - p_z/2$ and
\begin{eqnarray}
\overline{u}(p_1) \gamma_5 \gamma_\lambda u(p_1) = 2 S_\lambda~.
\end{eqnarray}
The function $F_5(z_+,z_-)$ is related to the
polarized distribution function $f_5(z_+,z_-)$ by
\begin{eqnarray}
\label{vecval}
F_5(z_+,z_-) = \int_0^1 \frac{d\lambda}{\lambda^2}
f_5\left(\frac{z_+}{\lambda},\frac{z_-}{\lambda}\right)
\theta(\lambda-|z_+|)\theta(\lambda-|z_-|)~.
\end{eqnarray}
We re--write the denominators by
\begin{eqnarray}
\label{eqDENO}
\frac{1}{Q^2_z+i\varepsilon} = - \frac{1}{qp_-}\frac{1}{(\vartheta
- 2\beta  + i \varepsilon)}~,
\end{eqnarray}
defining
\begin{eqnarray}
\beta =  \frac{x}{x_{\PP}} = \frac{q^2}{2q.p_-}~.
\end{eqnarray}
The conservation of the electromagnetic current is easily seen
\begin{eqnarray}
q^{\mu}T_{\mu\nu}(p_1,p_2,S,q)  =
q^\nu T_{\mu\nu}(p_1,p_2,S,q)~.
\end{eqnarray}
It follows because the contraction with $q^{\alpha}$ leads to 
Levi--Civita symbols being contract with the same 4--vector.
By
\begin{eqnarray}
\label{eqFF}
\widehat{F}(\vartheta,\eta) = \int Dz F(z_+,z_-) 
\delta(\vartheta - z_- -z_+/\eta) = \int_\vartheta^{-{\rm sign}(\vartheta)
/\eta} \frac{dz}{z} \widehat{f}(z,\eta)
\end{eqnarray}
we change to the variable $\vartheta$, the main momentum fraction in
the subsequent representation. Eq.~(\ref{eqFF}) is the pre--form of
the Wandzura--Wilczek integral~\cite{WW}. It emerges seeking the
representation of vector--valued distributions~(\ref{vecval}) in terms
of scalar distributions, cf.~\cite{BGR,BR}. In most of the applications
these integrals remain. An exception is the Callan--Gross relation,
see Refs.~\cite{BR,BRDIFF}, where all these integrals cancel and only
scalar distribution functions remain.
Here
the distribution function $\widehat{f}_5(z,\eta)$ is related to
$f_5(z_+,z_-)$ by
\begin{eqnarray}
\label{eqFF1}
\widehat{f}_5(z,\eta) = \int^{\eta(1-z)}_{\eta(1+z)} d\rho~
\theta(1-\rho) \theta(\rho+1) f_5(\rho,z-\rho/\eta)~,
\end{eqnarray}
with $\rho = z_+/\eta$.

The Compton amplitude takes the following form:
\begin{eqnarray}
\label{eqCOMP1}
T_{\mu\nu}^{\rm pol}(p_-,S,q) &=&
4i  \varepsilon_{\mu\nu\lambda\sigma} \int_{+1/\eta}^{-1/\eta}
d\vartheta \Biggl\{\frac{q^\lambda S^\sigma}{Q_z^2+i\varepsilon}
- q.S
\frac{\vartheta q^\lambda p_-^\sigma}
{(Q_z^2+i\varepsilon)^2}\Biggr\} \nonumber\\
& &~~~~~~~~~~~~~~~~~~~~~~~~~~~~~~~~~~~~~~~~~~~~~~~
\times \int_\vartheta^{-{\rm sign}(\vartheta)/\eta}\frac{dz}{z}
\widehat{f}_5(z,\eta)~.
\end{eqnarray}
The $\vartheta$--integral in Eq.~(\ref{eqCOMP1}) can be simplified
using the identities
\begin{eqnarray}
\int_{+1/\eta}^{-1/\eta} d\vartheta \frac{\vartheta^k}
{(\vartheta-2\beta+i\varepsilon)^2}\int_\vartheta^{{\rm sign}(\vartheta)
/\eta} \frac{dz}{z} \widehat{f}_5(z,\eta)
&=&
\int_{+1/\eta}^{-1/\eta} d\vartheta \frac{k \vartheta^{k-1}}
{(\vartheta-2\beta+i\varepsilon)}\int_\vartheta^{{\rm sign}(\vartheta)
/\eta} \frac{dz}{z} \widehat{f}_5(z,\eta) \nonumber\\    & &
~~~~~~~-\int_{+1/\eta}^{-1/\eta} d\vartheta \frac{\vartheta^{k-1}
\widehat{f}_5(\vartheta,\eta)}
{(\vartheta-2\beta+i\varepsilon)}~.
\end{eqnarray}
As we work in the approximation of $M^2, t << |q^2|$ 
the vector $p_-$ obeys the representation
\begin{eqnarray}
\label{vect}
p_- &=& - x_\PP p_1~.
\end{eqnarray}
Using these variables the Compton amplitude reads
\begin{eqnarray}
\label{eqCOMP2}
T_{\mu\nu}^{\rm pol}(p_1,S,q) &=&
-4i  \varepsilon_{\mu\nu\lambda\sigma} \int_{+1/\eta}^{-1/\eta}
\frac{d\vartheta}{\vartheta - 2 \beta + i \varepsilon} \Biggl\{
\left[\frac{q^\lambda S^\sigma}{q.p_1}
 + \frac{q.S}{(q.p_1)^2} q^\lambda p_1^\sigma
\right]
\int_{\vartheta}^{-{\rm sign}(\vartheta)/\eta} \frac{dz}{z}
\Hat{\Hat{f}}_5(z,\eta) \nonumber\\
& & ~~~~~~~~~~~~~~~~~~~~~~~~~~~~~~~~~~~~- \frac{q.S}{(q.p_1)^2}
q^\lambda p_1^\sigma
\Hat{\Hat{f}}_5(\vartheta,\eta) \Biggr\}~.
\end{eqnarray}
Here,
\begin{eqnarray}
\Hat{\Hat{f}}_5(z,\eta) =  \frac{1}{x_\PP} \widehat{f}_5(z,\eta)~.
\end{eqnarray}
Taking the absorptive part one obtains\footnote{The `imaginary part'
concerns that of the Schwartz-distribution Eq.~(\ref{eqDENO}).
Because of the relations, Eqs.~(\ref{eqP2C},\ref{eqP2B}) an overall $i$
emerges in the hadronic tensor.}
\begin{eqnarray}
\label{eqIMPAR}
W_{\mu\nu}^{\rm pol} &=& 
\frac{1}{2\pi}~{\sf Im}~T_{\mu\nu}^{\rm pol}(p_1,p_2,q) \nonumber\\
           &=&       
i \varepsilon_{\mu\nu\lambda\sigma} \frac{q^\lambda S_1^\sigma}{q.p_1}
{\sf G}_1(\beta,\eta,Q^2)
+ i  \varepsilon_{\mu\nu\lambda\sigma} \frac{q^\lambda (p_1.q S^\sigma
- S.q p_1^\sigma)}{(p_1.q)^2}
{\sf G}_2(\beta,\eta,Q^2)~,
\end{eqnarray}
where
\begin{alignat}{4}
\label{eqG1}
{\sf G}_1(\beta,\eta,Q^2) &=& \sum_{q=1}^{N_f}
e_q^2\left[\Delta f_q^D(\beta,Q^2,x_{\PP})+ \Delta
\overline{f}_q^D(\beta,Q^2,x_{\PP})
\right]
&\equiv&  g_1^{D(3)}(x,Q^2,x_{\PP})~, \\
\label{eqG2}
{\sf G}_2(\beta,\eta,Q^2) &=& 
-{\sf G}_1(\beta,\eta,Q^2) + \int_\beta^1 \frac{d\beta'}{\beta'}
 {\sf G}_1(\beta',\eta,Q^2)
&\equiv&  g_2^{D(3)}(x,Q^2,x_{\PP})~,
\end{alignat}
with $N_f$ the number of flavors, choosing the factorization scale
$\mu^2 = Q^2$. As we were working in the twist--2 approximation, the
Wandzura--Wilczek relation (\ref{eqG2}) describes
${\sf G}_2(\beta,\eta,Q^2)$.

To derive the representation for the diffractive parton densities
$\Delta f_q^D$, Eq.~(\ref{eqG1}), we consider the symmetry relation
for the polarized distribution functions $F^A(z_1,z_2)$, Ref.~\cite{BR},
\begin{eqnarray}
\label{eqSYM1}
F^A_5(z_1,z_2) =  F^A_5(-z_1,-z_2)~.
\end{eqnarray}
It translates into
\begin{eqnarray}
\label{eqSYM2}
\widehat{F}^A_5(\vartheta,\eta) =  \widehat{F}^A_5(-\vartheta,\eta)~,
\end{eqnarray}
and, cf. Eq.~(\ref{eqFF}),
\begin{eqnarray}
\label{eqSYM3}
\Hat{\Hat{f}}^A_5(\vartheta,\eta) =  \Hat{\Hat{f}}^A_5(-\vartheta,\eta)~.
\end{eqnarray}
The polarized diffractive quark and anti--quark densities are given by
\begin{eqnarray}
\label{eqPART}
\sum_{q=1}^{N_f} e_q^2 \Delta
f_q^D(\beta,Q^2,x_{\PP}) &=&
\Hat{\Hat{f}}_5(2\beta,\eta,Q^2) \nonumber\\
\sum_{q=1}^{N_f} e_q^2 \Delta
\overline{f}_q^D(\beta,Q^2,x_{\PP}) &=&
  \Hat{\Hat{f}}_5(-2\beta,\eta,Q^2)~.
\end{eqnarray}
Unlike in the deep--inelastic case, where the scaling variable
$x~\epsilon~[0,1]$, the support of the distributions
$\Delta f_q^D(\beta, Q^2, x_\PP)$ is $x~\epsilon~[0, x_\PP]$.

We express the diffractive parton densities in terms of
the distribution function $f_5(z_+,z_-)$ directly
\begin{eqnarray}
\label{eqREL}
\Delta
\Hat{\Hat{f}}_5(\pm 2\beta,\eta,Q^2)
 =  \frac{1}{x_\PP}
\int_{-\frac{x_{\PP} \pm 2x}{2-x_{\PP}}}
                              ^{-\frac{x_{\PP} \mp 2x}{2-x_{\PP}}}
d \rho f_5(\rho,\pm 2\beta + \rho(2-x_{\PP})/x_{\PP};Q^2)~.
\end{eqnarray}
The latter relations are needed to compare experimental quantities with
those which might be obtained measuring the corresponding operators
on the lattice.

Finally, we would like to make a remark on the evolution of the
diffractive parton densities being derived above. In a previous paper
\cite{BRDIFF} the corresponding evolution equations for unpolarized 
diffractive scattering have been derived in detail. Also here one
may start with the general formalism for non-forward scattering,
see e.g.~\cite{BGR}, and discuss the evolution of the scalar operators.
The evolution equations are independent of the parameter $\kappa_+$
emerging in the anomalous dimensions
$\gamma^{AB}_5(\kappa_+,\kappa_-,\kappa_+',\kappa_-';\mu^2)$ which 
therefore may be set to zero. Moreover, the all-order rescaling relation
\begin{eqnarray}
\label{eqresc}
\gamma^{AB}(\kappa_+,\kappa_-,\kappa_+',\kappa_-';\mu^2) = \sigma^{d_{AB}}
\gamma^{AB}(\sigma\kappa_+, \sigma\kappa_-,\sigma\kappa_+',
\sigma\kappa_-')~,
\end{eqnarray}
holds, with $d_{AB} = 2 + d_A - d_B$, $d_q=1, d_G=2$. A straightforward
calculation leads to the evolution equation for the polarized (singlet)
diffractive parton densities $f^A_5(\vartheta,\eta;\mu^2)$ in the
momentum fraction $\vartheta$
\begin{eqnarray}
\label{eqEV2}
\mu^2 \frac{d}{d\mu^2} f^A(\vartheta,\eta;\mu^2)
=  \int_\vartheta^{-{\rm sign}(\vartheta)/\eta}  \frac{d\vartheta'}
{\vartheta'} P^{AB}_5\left(\frac{\vartheta}{\vartheta'},\mu^2\right)
f_B(\vartheta',\eta;\mu^2)~.
\end{eqnarray}
The splitting functions $P_5^{AB}$ are the {\it forward} splitting 
functions~\cite{SPLIT}\footnote{For the non--forward anomalous dimensions
see \cite{SPNF}.}, which are independent of $\eta$ resp. 
$x_{\PP}$. Taking the 
absorptive part the usual evolution equations are obtained, with the
difference that the evolution takes place in the variable $\beta$.
The non-forwardness $\eta$ or $x_\PP$ behave as plain parameters.
\begin{eqnarray}
\label{eqEV4}
\mu^2 \frac{d}{d\mu^2} f^{D}_A(\beta,x_{\PP};\mu^2)
= \int_\beta^1 \frac{d\beta'}{\beta'} P_{5,A}^B
\left(\frac{\beta}{\beta'}; \mu^2\right)
f_B^D(\beta',x_{\PP};\mu^2)~.
\end{eqnarray}

We expressed the Compton amplitude with the help of the light--cone 
expansion at short distances and applied this representation to the
process of deep--inelastic diffractive scattering using Mueller's
generalized optical theorem. This representation is {\it not} limited to
leading twist operators but can be extended to all higher twist
operators. The corresponding evolution equations for
the higher twist hadronic matrix elements depend on more than one
momentum fraction $\vartheta_i$, which have a less trivial connection to
the outer kinematical variables similar to the case of deep--inelastic
scattering~\cite{BRRZ}. The construction is similar to the
above  and applies as well the generalized optical theorem. The evolution
of the associated parton correlation functions is for the same reason
forward.
\section{Conclusions}
\label{sec-5}

\vspace{1mm}
\noindent
The differential cross section of polarized deep--inelastic 
$ep$--diffractive scattering for pure photon exchange is described by 
eight structure functions. They depend on the four kinematic variables, 
$x, Q^2, x_{\PP}$ and $t$. In the limit of small values of $t$ and
neglecting target
masses two structure functions contribute. In the generalized Bjorken
range and the presence of a sufficiently large rapidity gap the scaling 
violations of hard diffractive scattering can be described within 
perturbative QCD. In this range processes, which are dominated by
light--cone contributions, are described. The scattering amplitude can be
rewritten using
Mueller's generalized optical theorem moving the outgoing diffractive
proton into an incoming anti-proton. In this kinematical domain
diffractive scattering {\it is} deep--inelastic scattering off a state
$\langle p_1,S_1,-p_2|$. Non-forward techniques may be used to describe
this process. In this way the two--variable polarized amplitudes
turn into the polarized diffractive parton densities, which depend
on one momentum fraction and a parameter $\eta$, which describes
the non--forwardness, and is directly related to the variable $x_\PP$.
For the absorptive part the scaling variable can be expressed by
the variable $\beta$, which also is the variable on which the
evolution kernels act in the twist--2 contributions, whereas $x_\PP$ 
remains as a simple parameter of the process. In the limit 
$t, M^2 \rightarrow 0$ the twist--2 contributions to the two structure 
functions $g_{1,2}^{D(3)}(x,Q^2,x_\PP)$ are related by a 
Wandzura--Wilczek relation in the variable $\beta = x/x_\PP$.
The approach followed in the present paper for twist--2 operators
can be synonymously extended to higher twist--operators in the kinematic
domain of the general Bjorken limit.

\vspace{2mm}
\noindent
{\bf Acknowledgement.}~For discussions we would like to thank J.~Eilers 
B.~Geyer, and X.~Ji. We thank J.~Dainton, M.~Erdmann, and D.~Wegener  for their
interest in the present work.

\end{document}